# Wafer-Scale Demonstration of High-Voltage β-$Ga_2O_3$ MOSFETs with Excellent Uniformity and >3 kV Breakdown Voltages

Ningtao Liu[1,2,5]*, Hengrui Zhang[2,5], Shujun Zhu[2,5], Zhihao Yan[2], Dongyang Han[2], Shen Hu[3], Li Ji[3,]*, Ning Xia[4], Jichun Ye[1,2,]* and Wenrui Zhang[1,2,]*

1 Yongjiang Laboratory, Ningbo 315201, China.

2 Ningbo Institute of Materials Technology and Engineering, Chinese Academy of Sciences, Ningbo 315201, China.

3 School of Microelectronics, Fudan University, Shanghai, 200433, China.

4 Hangzhou Garen Semiconductor Co., Ltd., Hangzhou, Zhejiang 311200, China.

[5]Ningtao Liu, Hengrui Zhang and Shujun Zhu contributed equally to this work.

Corresponding author: Ningtao Liu, Li Ji, Jichun Ye and Wenrui Zhang,

E-mail: ningtao-liu@ylab.ac.cn, lji@fudan.edu.cn, jichun-ye@ylab.ac.cn, wenrui-zhang@ylab.ac.cn

## Abstract

This study demonstrates a wafer-scale growth of a 2-inch Si-doped β-$Ga_2O_3$ (100) epitaxial wafer and the realization of uniform, high-voltage lateral β-$Ga_2O_3$ MOSFET arrays. The 2-inch homoepitaxial β-$Ga_2O_3$ (100) film grown by MOCVD exhibit excellent crystalline uniformity with an average rocking curve FWHM of ~27.0 arcsec and a low surface roughness less than 1 nm, alongside a uniform net doping concentration on the value of $4.60\times10^{17}$ $cm^{-3}$. The fabricated MOSFETs deliver a threshold voltage of -31.75 V, a drain-current on/off ratio $>10^9$, a specific on-resistance of 126.52 $m\Omega\cdot cm^2$ and breakdown voltage exceeding 3 kV. Statistical analysis across the entire wafer presents good device uniformity, with threshold voltages ranging from -28 V to -36 V, output current densities of 60-75 mA/mm, and a breakdown voltage over 3 kV. These results provide the demonstration using the 2-inch β-$Ga_2O_3$ epitaxial wafer to realize high-voltage β-$Ga_2O_3$ MOSFETs with wafer-scale performance uniformity for next-generation power device application.

**Keywords:** 2-inch $Ga_2O_3$ Homoepitaxial wafer, MOCVD, MOSFET, Breakdown voltage, Statistical analysis.

## I. Introduction

The ultrawide bandgap semiconductor beta phase gallium oxide (β-$Ga_2O_3$) has emerged as a promising candidate for next-generation power electronics due to its exceptional material properties, including a wide bandgap (~4.9 eV) and a high critical breakdown field (>8 MV/cm) [1], [2], [3]. Recent progress in cost-effective melt-growth techniques has enabled the production of high-quality, large-area single-crystal β-$Ga_2O_3$ substrates, offering a scalable pathway toward wafer-based manufacturing and facilitating the material's industrial application [3], [4]. A pivotal challenge for the commercial adoption of β-$Ga_2O_3$ lies in transitioning from proof-of-concept devices on small-area samples to reproducible, large-area fabrication on uniform epitaxial wafers. The performance, yield, and reliability of power transistors are intrinsically linked to the homogeneity of the underlying material, which includes its crystalline quality, surface morphology, and doping uniformity across the entire wafer [5]. Deviations from these parameters can result in substantial inter-device variation in critical device metrics, including threshold voltage, on-state current, and breakdown strength. This, in turn, has the potential to exert an adverse effect on the performance, reliability and manufacturability of the power circuit or module [6], [7]. Consequently, comprehensive wafer-scale characterization and statistical analysis of device performance are essential to assess the technological readiness of β-$Ga_2O_3$.

While individual device records are important to justify the advantage of β-$Ga_2O_3$ power devices, achieving consistently high performance across a full wafer now becomes more stringent for pursuing next-step commercial application. Previous reports have mostly focused on metrics for a single MOSFET unit within specific regions of a small sample (10×10 $mm^2$ or less), with less emphasis on the statistical distribution of a number of MOSFET devices that dictate the process yield [7], [8], [9]. Construction high-performance MOSFET arrays on inch-scale wafers remains challenging due to the need for uniform homoepitaxial growth and wafer-scale device processing [9]. In this work, we demonstrate high-uniformity lateral MOSFETs on a 2-inch β-$Ga_2O_3$ epitaxial wafer. The 2-inch β-$Ga_2O_3$ (100) film is grown by Metal-Organic Chemical Vapor Deposition (MOCVD), which exhibits excellent homogeneity in crystallinity, surface morphology, and doping profile. The lateral β-$Ga_2O_3$ MOSFETs achieve a high on/off ratio (>$10^9$), a low specific on-resistance ($R_{on,sp}$ = 126.52 mΩ·$cm^2$), and a breakdown voltage ($V_{br}$) >3 kV. Full-wafer statistics confirm tight distributions in threshold voltage and output current density, with all devices uniformly exceeding an on/off ratio of $10^9$ and a $V_{br}$ over 3 kV, thereby indicating outstanding wafer scale performance and uniformity for power applications.

**II. Experiments**

The homoepitaxial films were growth on 2-inch Fe-doped (100)-orientation β-$Ga_2O_3$ substrates (Hangzhou Garen Semiconductor Co., Ltd.) with a 4° off-cut toward the [00-1] orientation by MOCVD. Prior to growth, the 2-inch substrates were sequentially cleaned by acetone, ethanol, and deionized water. Subsequently, the wafer was loaded into chamber and subjected to an in-situ annealing at 750 °C for 5 minutes to further improve the surface condition. The 2-inch wafer-level homoepitaxial films were grown at 750 °C and 80 mbar. Triethylgallium (TEGa) and high-purity oxygen gas were used as gallium and oxygen precursors, respectively. The n-type doping was achieved using dilute silane ($SiH_4$) as the precursor. MOSFETs without terminal structures was fabricated based on 2-inch epitaxial wafer. To achieve electrical isolation between devices, inductive coupled plasma (ICP) dry etching was employed to remove the $Ga_2O_3$ epitaxial layer outside the active regions, forming isolation platforms. The etching depth was approximately 500 nm using a $BCl_3$/$Cl_2$/Ar mixed gas. Source (S) and drain (D) electrode regions were defined via photolithography using a mask for proximity exposure. Residual unexposed photoresist was removed by oxygen plasma pretreatment prior to SD electrode growth. Ti/Au (30/50 nm) composite electrodes were patterned using lift-off techniques after electron beam evaporation deposition. An $Al_2O_3$ insulating layer (50 nm thick) was grown via thermal deposition at 200°C using plasma-enhanced atomic layer deposition (PEALD). A plasma flow of $O_2$ excited by trimethylaluminum and $O_2$ ensured precise control over layer growth. The gate electrode metal (Ni/Au = 30/50 nm) was deposited by magnetron sputtering to complete the final fabrication of the device. The transfer, output and three-terminal breakdown characteristics were evaluated using an Keysight B1500 Device Analyzer.

## III. Results and discussion

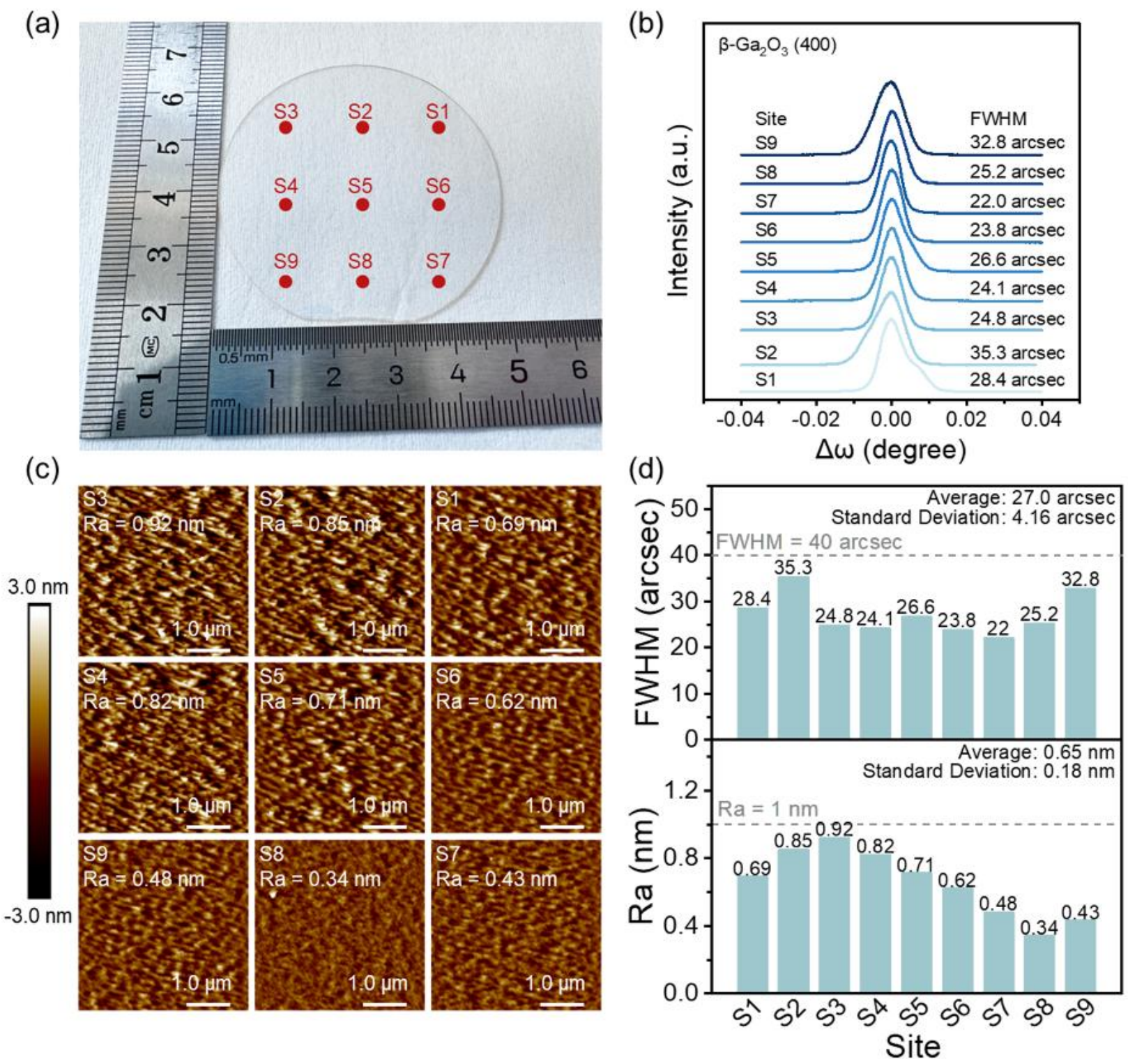

**Fig. 1.** (a) Photograph of a 2-inch β-$Ga_2O_3$ (100) epitaxial wafer. The crystallinity and surface morphology uniformity is evaluated based on the nine representative locations (S1-S9) distributed evenly over the wafer. (b) The measured X-ray rocking curves of the β-$Ga_2O_3$ (400) diffraction plane and (c) AFM images of the S1-S9 locations of the 2-inch epitaxial wafer. (d) A uniformity summary of FWHM and surface roughness of the S1-S9 locations showing excellent crystallinity and low roughness of the 2-inch β-$Ga_2O_3$ epitaxial wafer.

To evaluate the wafer-scale uniformity of the 2-inch epitaxial wafer, nine representative regions are selected for systematic structural and electrical characterization, as depicted in **Fig. 1(a)**. **Fig. 1(b)** presents the X-ray rocking curves measured across these regions, which show a consistent full width at half maximum (FWHM) of approximately 27.0 arcsec with a standard deviation of 4.16 arcsec, indicating high crystalline quality over the entire wafer. In addition, low surface roughness is critical for the fabrication of high-performance lateral transistors [1], [3]. Therefore, atomic force microscopy (AFM) is employed to evaluate the surface morphology. As shown in **Figs. 1(c)** and **(d)**, all examined regions exhibit smooth surfaces with an average root-mean-square roughness of 0.65 nm.

As demonstrated in **Fig. 2(a)**, the net doping concentration ($N_D$-$N_A$) of a representative S8 region of the 2-inch epitaxial wafer, obtained from mercury-probe C–V measurements at 100 kHz, is approximately 4.60 × $10^{17}$ $cm^{-3}$. **Figs. 2(b)** and **(c)** present the corresponding depth profile of $N_D$-$N_A$ derived from C-V measurements, alongside the statistical distribution of carrier concentrations across various wafer regions. The results indicate that the $N_D$-$N_A$ value is 4.60 ± 0.45 × $10^{17}$ $cm^{-3}$, showing decent uniformity across the 2-inch wafer. As shown in **Fig. 2(d)**, the thickness of the β-$Ga_2O_3$ epitaxial layer ranges from 209.22 nm to 213.49 nm, with an average thickness of 211.68 nm and a standard deviation of only 1.14 nm, corresponding to a total thickness variation of merely ~2.0%, confirming excellent thickness uniformity.

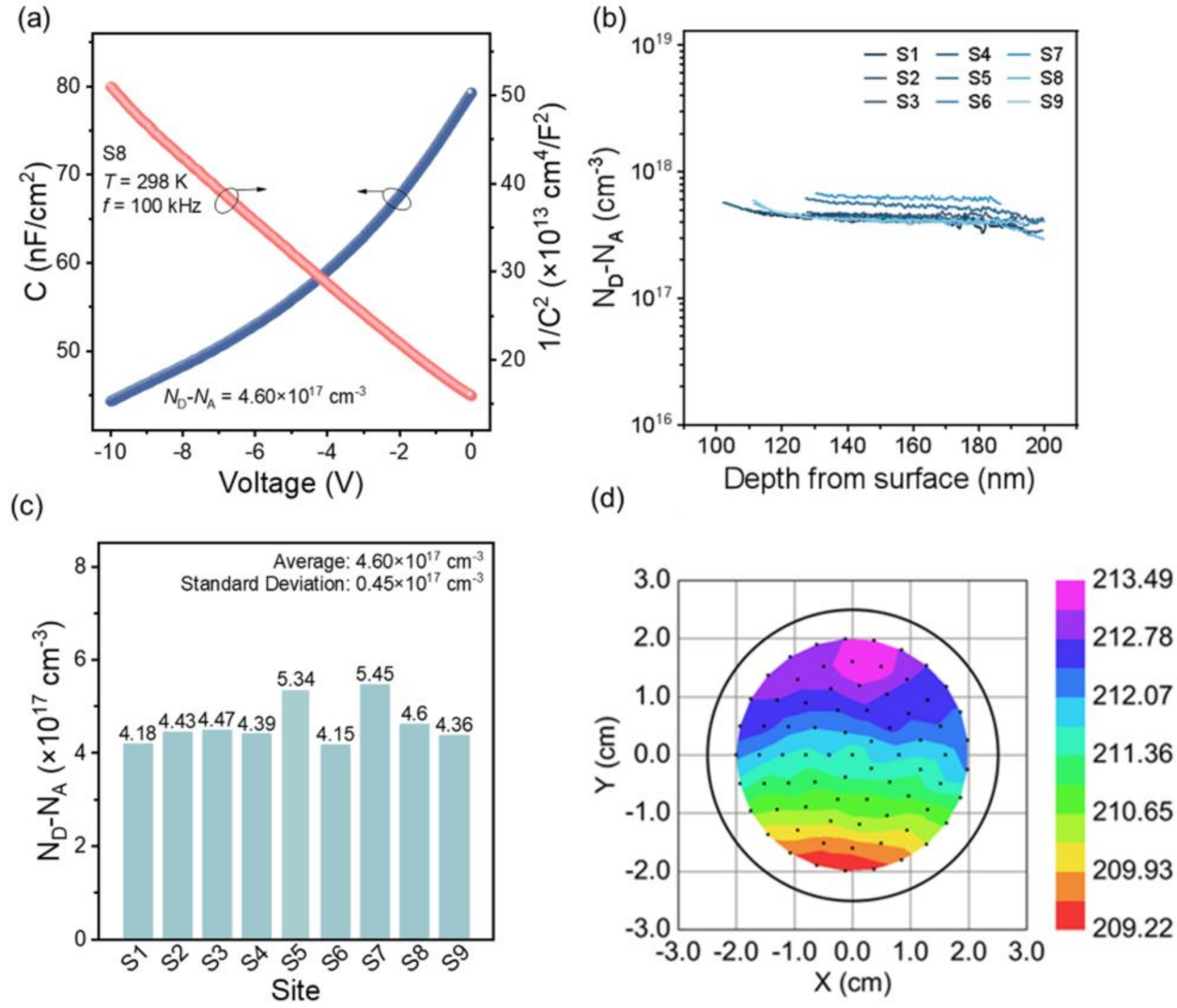

**Fig. 2.** (a) The measured C-V characteristics of a representative location of the epitaxial wafer for determining the carrier concentration. (b) The depth distribution of the carrier concentration (the illustration shows the internal structure of the mercury probe) and (c) a statistic summary of nine regions from the two-inch epitaxial wafer demonstrating the uniform carrier concentration. (d) Thickness mapping of the β-$Ga_2O_3$ epitaxial layer on a 2-inch sapphire control wafer.

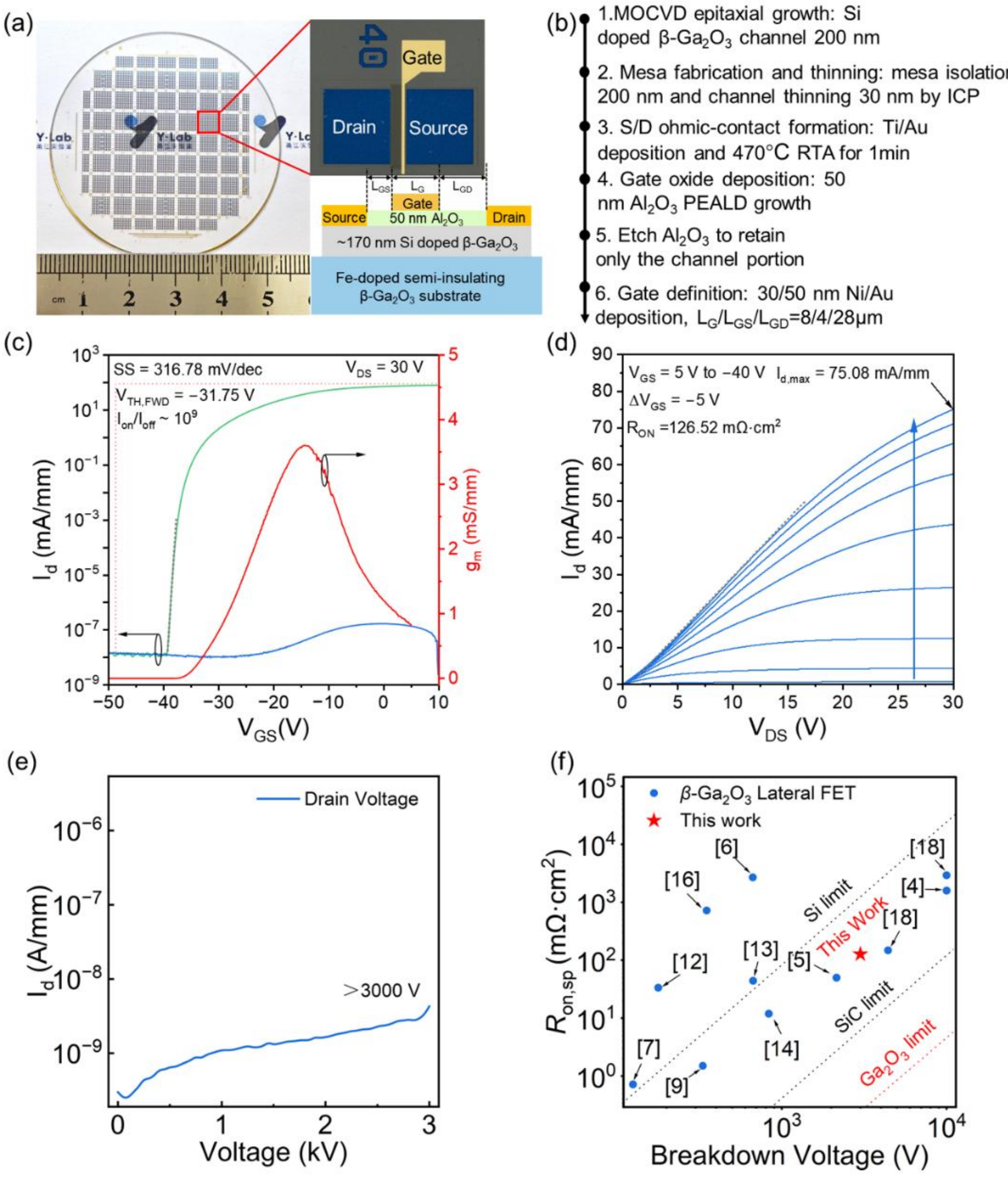

**Fig. 3**. (a) The optical image of a MOSFET array based on a fully processed 2-inch β-$Ga_2O_3$ wafer, along with a magnified view of an individual MOSFET and cross-sectional schematic of the β-$Ga_2O_3$ MOSFET device. (b) The process flow of the β-$Ga_2O_3$ MOSFET. (c) The measured transfer, (d) output and (e) breakdown characteristics for a representative MOSFET device. (f) The benchmark of $R_{on,sp}$ vs. $V_{br}$ of the high-voltage β-$Ga_2O_3$ MOSFETs in this work and the previously reported β-$Ga_2O_3$ MOSFETs.

The fully processed MOSFET arrays based on 2-inch β-$Ga_2O_3$ wafer, along with the schematic cross-section and magnified view of an individual device, are shown in **Fig. 3(a)**. The fabrication process flow of the $Ga_2O_3$ MOSFET devices is illustrated in **Fig. 3(b)**. The transfer and output characteristics of a device with a gate-to-drain length ($L_{GD}$) of 28 μm are presented in **Figs. 3(c)** and **(d)**, respectively. As seen, the threshold voltage of the device is around -31.75 V, and the drain current on-off ratio of $>10^9$ and subthreshold slope (SS) of ~316.78 mV/dec. The high SS value is likely due to the absence of surface pretreatments prior to gate oxide deposition [7]. The output characteristics of the single device show a maximum drain density of 75.08 mA/mm and $R_{on.sp}$ of 126.52 mΩ·cm$^2$ at a gate voltage ($V_{GS}$) of 5 V, which is comparable to other reports [4], [5], [6], [7], [9], [12], [13], [14], [16], [18]. **Fig. 3(e)** shows the three-terminal off-state breakdown characteristics at $V_{GS}$ = - 40 V. During the measurement, the device is immersed in a fluorinert solution in order to avoid premature breakdown in air. The measured $V_{br}$ exceeds 3 kV, alongside an average breakdown field of approximately 1 MV/cm, which is consistent with previously reported results for similar devices (**Fig. 3(f)**) [4], [5], [6], [7], [9], [12], [13], [14], [16], [18].

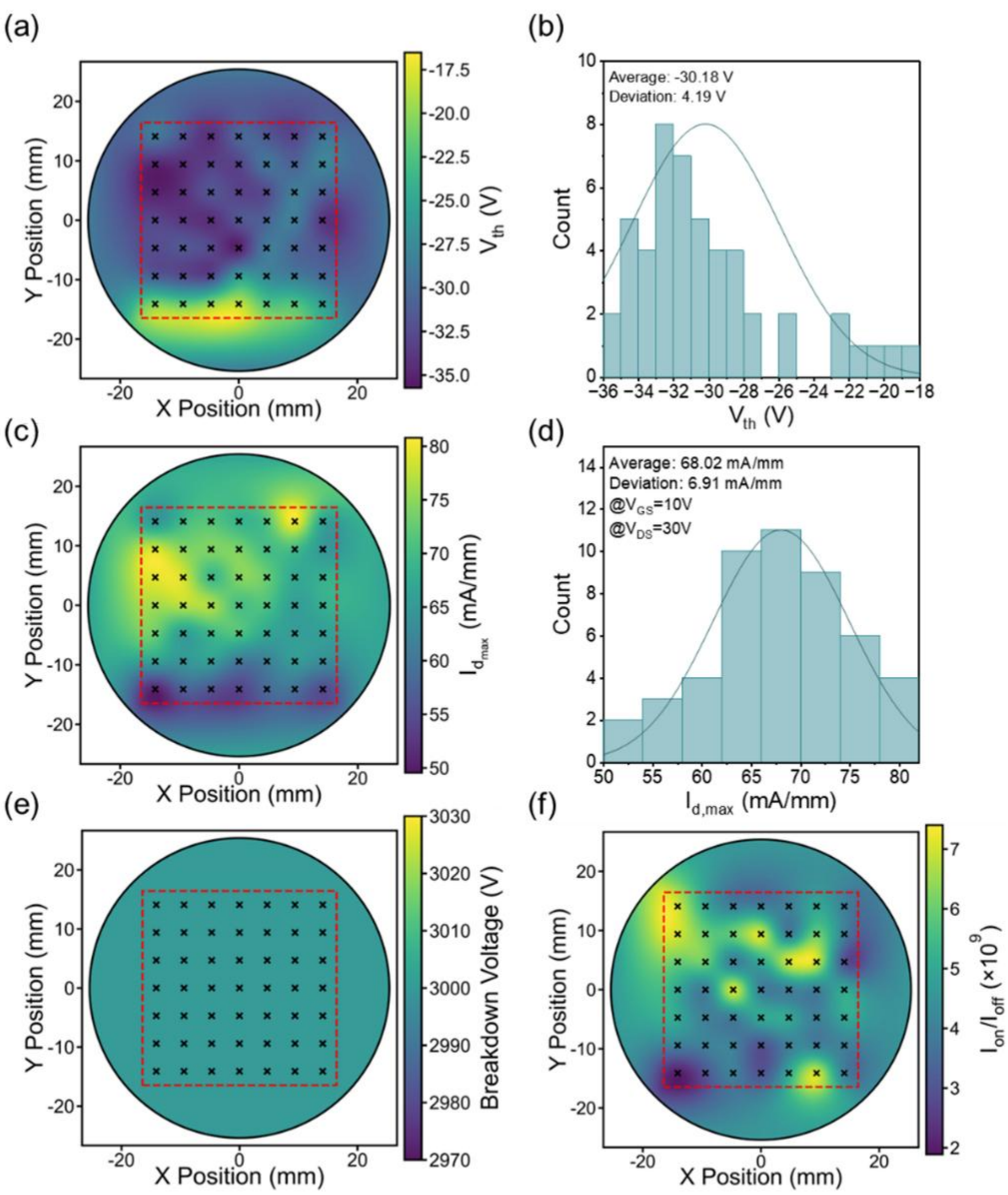

**Fig 4.** Spatially resolved mappings and corresponding statistical distributions of (a, b) the measured threshold voltage, (c, d) output current density, and (e, f) Spatially resolved mappings distributions of the measured $V_{br}$ and drain-current on/off ratio for MOSFETs across the 2-inch homoepitaxial

wafer, respectively.

To statistically evaluate the electrical performance of the fabricated devices across the 2-inch epitaxial wafer, all devices on the wafer were systematically characterized under the same conditions as in **Figs. 3(c)–(d)**. The results are summarized in **Fig. 4**. Across the 2-inch wafer, the threshold voltage ($V_{th}$) for 85% of the devices falls within a narrow range of -28 V to -36 V, corresponding to a variation of approximately 4 V. Similarly, 71% of the devices exhibit an output current density between 60 mA/mm and 75 mA/mm, with a variation of about 6.9 mA/mm. Moreover, as illustrated in **Figs. 4(e)** and **(f)**, all devices fabricated on the 2-inch epitaxial wafer demonstrate a drain-current on-off ratio exceeding $10^9$ and a $V_{br}$ greater than 3 kV, outperforming previously reported values [13], [18]. The above statistical results demonstrate the excellent performance and uniformity of β-$Ga_2O_3$ for power transistor applications.

## IV. Conclusion

In summary, a high-quality, 2-inch β-$Ga_2O_3$ epitaxial wafer is grown via MOCVD, exhibiting uniform structural and net doping concentration on the value of $4.60\times10^{17}$ $cm^{-3}$. Lateral MOSFET devices fabricated on the wafer achieve a high on/off ratio (>$10^9$), a low $R_{on,sp}$ (126.52 mΩ·$cm^2$), and a $V_{br}$ >3 kV. Full-wafer statistics confirm tight distributions in threshold voltage and output current density, with all devices uniformly exceeding an on/off ratio of $10^9$ and a $V_{br}$ of 3 kV, showcasing outstanding wafer -scale performance and uniformity for power applications.

## CRediT authorship contribution statement

**Ningtao Liu:** Writing – original draft, review & editing, Investigation, Funding acquisition, Formal analysis, Conceptualization. **Hengrui Zhang:** Investigation, Data curation. **Shujun Zhu:** Investigation, Data curation. **Zhihao Yan:** Data curation. **Dongyang Han:** Investigation. **Shen Hu:** Resources. **Li Ji:** Resources,Writing – review & editing. **Ning Xia:** Resources. **Jichun Ye:** Writing – review & editing, Resources, Project administration. **Wenrui Zhang:** Writing – review & editing,

Supervision, Funding acquisition, Formal analysis, Conceptualization.

## Declaration of competing interest

The authors declare that they have no known competing financial interests or personal relationships that could have appeared to influence the work reported in this paper.

## Acknowledgment

This work was supported by the National Natural Science Foundation of China (Grant No. 62204244) and the Zhejiang Provincial Natural Science Foundation of China (Grant No. LQ23F040003). Part of the research was supported by Ningbo Yongjiang Talent Introduction Programme (Grant No. 2021A-046-C).